\documentclass[prd,longbibliography,superscriptaddress,showpacs,showkeys,twocolumn,floatfix,eqsecnum]{revtex4-1}
\usepackage{gensymb}
\usepackage{textcomp}
\usepackage{eurosym}
\usepackage{amsfonts}
\usepackage{url}
\usepackage{array}
\usepackage{amsthm}
\usepackage{bm}
\usepackage{palatino}
\usepackage[colorinlistoftodos]{todonotes}
\usepackage{mathpazo}
\usepackage{supertabular}
\usepackage{subfig}
\usepackage{amssymb}
\usepackage{eurosym}
\usepackage{amsmath}
\usepackage{graphics}
\usepackage{color}
\usepackage{graphicx}
\usepackage[font={footnotesize,it}]{caption}
\usepackage[colorlinks=true,
            linkcolor=blue,
            urlcolor=blue,
            citecolor=green]{hyperref}
\usepackage{tabstackengine}
\usepackage{graphicx}
\usepackage[english]{babel}
\usepackage{amsfonts}
\usepackage{eurosym}
\usepackage{calrsfs}
\usepackage{color}
\allowdisplaybreaks
 \newcommand{\bq}{\begin{equation}}
 \newcommand{\eq}{\end{equation}}
 \newcommand{\bqn}{\begin{eqnarray}}
 \newcommand{\eqn}{\end{eqnarray}}

\begin{document}

%Title of paper
\title {Thin accretion disk in the Simpson-Visser black-bounce and wormhole spacetime}

\author{Parth Bambhaniya}
\email{grcollapse@gmail.com}
\affiliation{International Center for Cosmology, Charusat University, Anand, GUJ 388421, India}
\author{Saurabh}
\email{sbhkmr1999@gmail.com}
\affiliation{P. D. Patel Institute of Applied Sciences, Charusat University, Anand, 388421, Gujarat, India}

\author{Kimet Jusufi}
\email{kimet.jusufi@unite.edu.mk}
\affiliation{Physics Department, State University of Tetovo, Ilinden Street nn, 1200,
Tetovo, North Macedonia}

\author{Pankaj S. Joshi}
\email{psjcosmos@gmail.com}
\affiliation{International Center for Cosmology, Charusat University, Anand, GUJ 388421, India}

\date{\today}

\begin{abstract}
 We compare the optical appearance of a thin accretion disk in the Simpson-Visser spacetime to the Schwarzschild black hole case in this paper. We calculate and illustrate the red-shift and observed flux distributions as viewed by distant observers at various inclination angles. Simpson-Visser family of metrics create Novikov-Thorne (NT) accretion disks images that nearly look like a Schwarzschild black hole's NT accretion disk. We have studied also the embedding diagram, the electromagnetic properties of the accretion disk such as the temperature and the radiation flux of the energy by the accretion disk and the accretion efficiency. Compared to the Schwarzschild black hole, we find that the temperature, radiation flux of the energy, and the luminosity of the accretion disk increase by increasing the regularization parameter $'l'$. We conclude that, based on astrophysical observational signatures in the properties of the electromagnetic spectrum, we can distinguish the wormhole geometries from the regular black holes (black-bounce) and the Schwarzschild black hole. 
\end{abstract}
 
\maketitle

\section{Introduction}
Black holes are arguably the most fascinating predictions of Einstein's general theory of relativity. It is now evident that both astronomical observations and the theoretical formulation of general relativity suggest their real existence in nature. The most striking evidence for their existence is the detection of the
gravitational waves \cite{gw} and the shadow image of the supermassive M87*  \cite{eh1,eh2}. Black holes are extremely important in explaining many other problems in astrophysics that have to do with the high-energy phenomena in the form of X-ray emission and jets, accretion of matter, quasi-periodic oscillations, and the motions of nearby objects in orbit around the hidden mass \cite{GRAVITY:2020gka}.

Despite all these successes, there are still some conceptual problems related to black holes in general relativity. More specifically, the main problem associated with black holes is the existence of singularities at the coordinate center. As of today, this problem has not been solved. Many physicists have speculated that a quantum theory of gravity can solve this problem. However, the phenomenology of black holes remains a hot topic after the EHT findings. The black hole mimickers such as regular black holes, wormholes, naked singularities, grava-stars, etc. are also growing significantly in the literature \cite{Hayward:2005gi,Visser:2003ge,joshi,Gao:2016bin,Mosani:2020ena,Mosani:2020mro,Mosani:2021,az,amir}.

The concept of a regular black hole was first introduced by Bardeen in 1968 \cite{Bardeen:1968}, see also the more recent references \cite{Bergmann-Roman,Hayward:2005,Bardeen:2014,Frolov:2014,Frolov:2014,Frolov:2014b,Frolov:2016,Frolov:2017,Frolov:2018,Cano:2018,Bardeen:2018,regular,beyond}, and it has been intuitively interesting due to its non-singular structure.  One motivation for exploring such compact objects is that they could be black hole alternatives. Furthermore, as is widely known, many of the possible observational signatures of the black holes such as precession of timelike bound orbits, gravitational lensing and shadow properties can be mimicked by black hole mimickers \cite{Bambhaniya:2019pbr,Joshi:2019rdo,Bambhaniya:2020zno,Bambhaniya:2021ybs,Nampalliwar:2021tyz,Jusufi:2020zln,Saurabh:2020zqg,Nampalliwar:2021oqr,Joshi:2020tlq,Dey:2019fpv}.

From a historical perspective, it's worth noting that Einstein and Rosen proposed the existence of "bridges" through space-time \cite{ERb} using the theory of general relativity. These bridges connect two places in space-time, allowing for the creation of Einstein-Rosen bridges, also known as wormholes. Such wormholes have been proved to be non-traversable. The notion of having traversable wormholes was later investigated in the pioneering work of Morris and Throne \cite{Morris}.

The exotic matter that satisfies the flare-out criterion and violates the weak energy condition is required to sustain the structure of a wormhole (see \cite{Morris:1988cz, Morris:1988tu}). However, quantum matter fields have recently been proved to provide enough negative energy to allow some wormholes to be traversed. As a result, an exotic matter with negative energy density and large negative pressure, with a greater value than the energy density is required to form such a traversable wormhole \cite{k1,k2}.

We can therefore try to test black holes and wormholes using astrophysical data. In Ref. \cite{k3} authors studied the possibility to test wormholes geometries in our galaxy using infalling and radiating gas accretion model and the motion of the S2 star. The possibility of testing wormholes by observing the galactic center using the infalling gas model was studied in Refs. \cite{b1,b2}. Other interesting works concerning the formation and the stability of wormholes can be found in Refs. \cite{s1,s2,s3}. Based on astrophysical observations, today we know that in the galactic center of many galaxies a disk is formed. Namely, such a disk is made of rapidly rotating gas that slowly spirals onto a compact body which is assumed to be a black hole. During such accretion, the gravitational energy through the friction of the heat converts into radiation, which partially escapes, and cools down the
accretion disk. The only information that we have about
accretion disk physics comes from this radiation, when it
reaches radio, optical and X-ray telescopes, allowing astronomers to analyze its electromagnetic spectrum, and
its time variability. The radiation
properties of thin accretion disks were further analyzed
in the recent papers \cite{Shaikh:2019hbm,1,2,3,4,5,6}, where the effects of photon capture by the black hole on the spin evolution were presented as well.  

Simpson and Visser proposed a very simple theoretically appealing spherically symmetric and static spacetime family \cite{Simpson:2018tsi}, derived from Schwarzschild geometry, that allows for a unique description of regular black hole and wormholes by smooth interpolation between these two possibilities using a length-scale parameter $'l'$ that drives the regularization of the central singularity. Its rotational form was recently shown as well \cite{Mazza:2021rgq}. In the present work, we are interested to use the Simpson and Visser metric and explore the NT accretion disk properties and images. It will be interesting to see whether or not we can distinguish the regular/non-regular black hole from the wormhole case.  This metric is of particular interest since it can describe a regular/non-regular black hole or traversable wormhole depending on the value of the regularization parameter $l$. Note that many physical properties of this metric, including the effect of charge, gravitational lensing, quasi-normal modes, and other effects were studied by many authors \cite{sv1,sv2,sv3,sv4,sv5,sv6,sv7,sv8}. In the present work, we are interested to study the optical appearance and the electromagnetic properties of the NT accretion disk around the Simpson-Visser metric. 

This paper is organized as follows. In Section II, we present the Simpson-Visser metric. In Section III, we study the embedding diagram. In Section IV, we study the NT accretion disk and the images. In Section V, we explore the electromagnetic properties of the accretion disk. Finally, in Section VI, we comment on our results. Throughout this paper, we consider gravitational constant $(G)$ and speed of light $(c)$ as unity.

\section{Simpson-Visser spacetime}
The proposed spherically symmetric and static Simpson-Visser metric specifies a class of black hole mimickers with a minimal surface in place of the central singularity. The geometry of spacetime is given by \cite{Simpson:2018tsi}, 
\begin{equation}
ds^2=-f(r)dt^2+\frac{dr^2}{f(r)}+(r^2+l^2)(d\theta^2+\sin^2\theta d\phi^2).
\label{SVmetric}
\end{equation}
where, 
\begin{equation}
    f(r)=\left(1-\frac{2M}{\sqrt{r^2+l^2}}\right).\; \; \; \;
 %   d\Omega^2=d\theta^2+\sin^2\theta d\phi^2.
\end{equation}
In the above spacetime, $M\ge 0$ represents the ADM mass, and $l>0$ is a parameter responsible for the regularisation of the central singularity and
possibly reflecting the quantum gravity effects. The above spacetime (\ref{SVmetric}) neatly interpolates between the standard Schwarzschild black hole and the Morris–Thorne traversable wormhole; Passing through a black-bounce (into a future incarnation of the universe), an extremal null-bounce (into a future incarnation of the universe), and a traversable wormhole are all intermediary stages. Therefore, this spacetime consist family of solutions as the black hole mimickers, which can be described as,
\begin{enumerate}
\itemsep-1pt
\item 
The ordinary Schwarzschild spacetime ($l=0$); 
\item
A ``black-bounce'' (regular black hole) with a one-way space-like throat  ($l<2M$);
\item
A one-way wormhole with a null throat ($l=2M$), 
\item
A traversable wormhole in the Morris--Thorne sense ($l>2M$) 
\end{enumerate} 
The Simpson and Visser metric is everywhere regular as long as the parameter $l$ is non-zero, resulting in an unusual kind of "regular black hole," where the"origin" at r = 0 can be space-like, null, or timelike. Moreover, the Carter–Penrose diagrams and curvature tensors are defined in \cite{Simpson:2018tsi} and shown that the Einstein tensor has non-zero (mixed) components. In the present work we are interested to understand the effect of $l$ on the physical properties of the electromagnetic waves emitted by the accretion disk. 
%By applying the Newman-Janis Algorithm (NJA) one can construct a rotating generalisation of a seed metric. Using Simpson-Visser metric as a seed metric, we can obtain a stationary, axially symmetric spacetime and that depends on mass (M), spin (a) and minimal surface of radius parameter $(l)$. The rotating Simpson-Visser spacetime metric is given by \cite{Simpson:2018tsi},
%\begin{eqnarray}
%    ds^2=-\left(1-\frac{2M\sqrt{r^2+l^2}}{\Sigma}\right) dt^2 + \frac{\Sigma}{\Delta} dr^2 + \Sigma d\theta^2 - \frac{4Masin^2\theta\sqrt{r^2+l^2}}{\Sigma} dt d\phi \nonumber\\
%    + \frac{\left((r^2+l^2+a^2)^2-\Delta a^2sin^2\theta \right)}{\Sigma}sin^2\theta d\phi^2.
%end{eqnarray}
	
\begin{figure}
    \centering
    \includegraphics[scale=0.79]{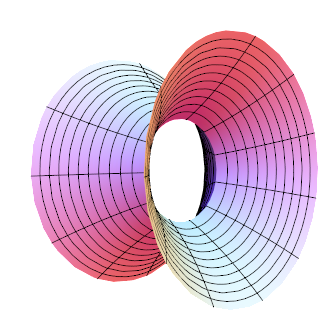}
    \caption{The geometry of Simpson-Visser wormhole embedded in a three-dimensional Euclidean space. We have used $l=2.1$ and $M=1$.}
    \label{geometry}
\end{figure}

\begin{figure*}
    \centering
    \includegraphics[scale=0.69]{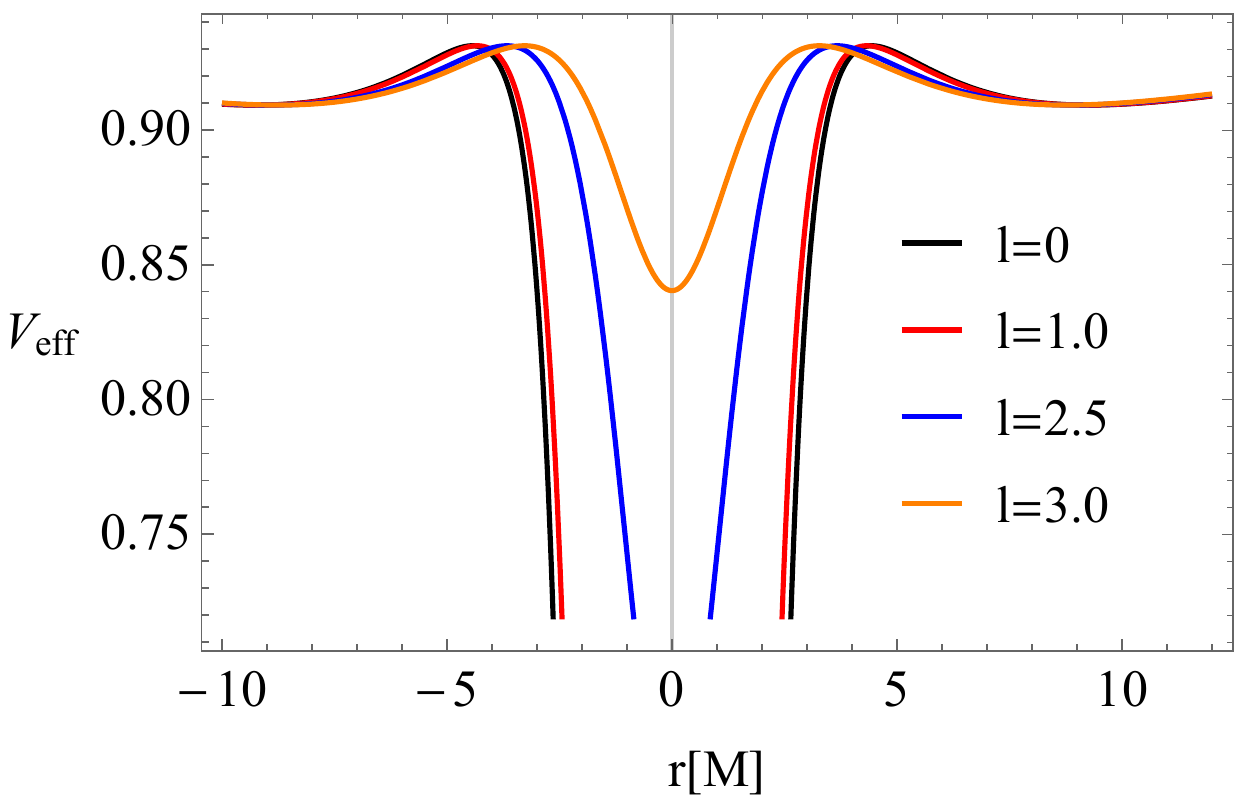}
        \includegraphics[scale=0.69]{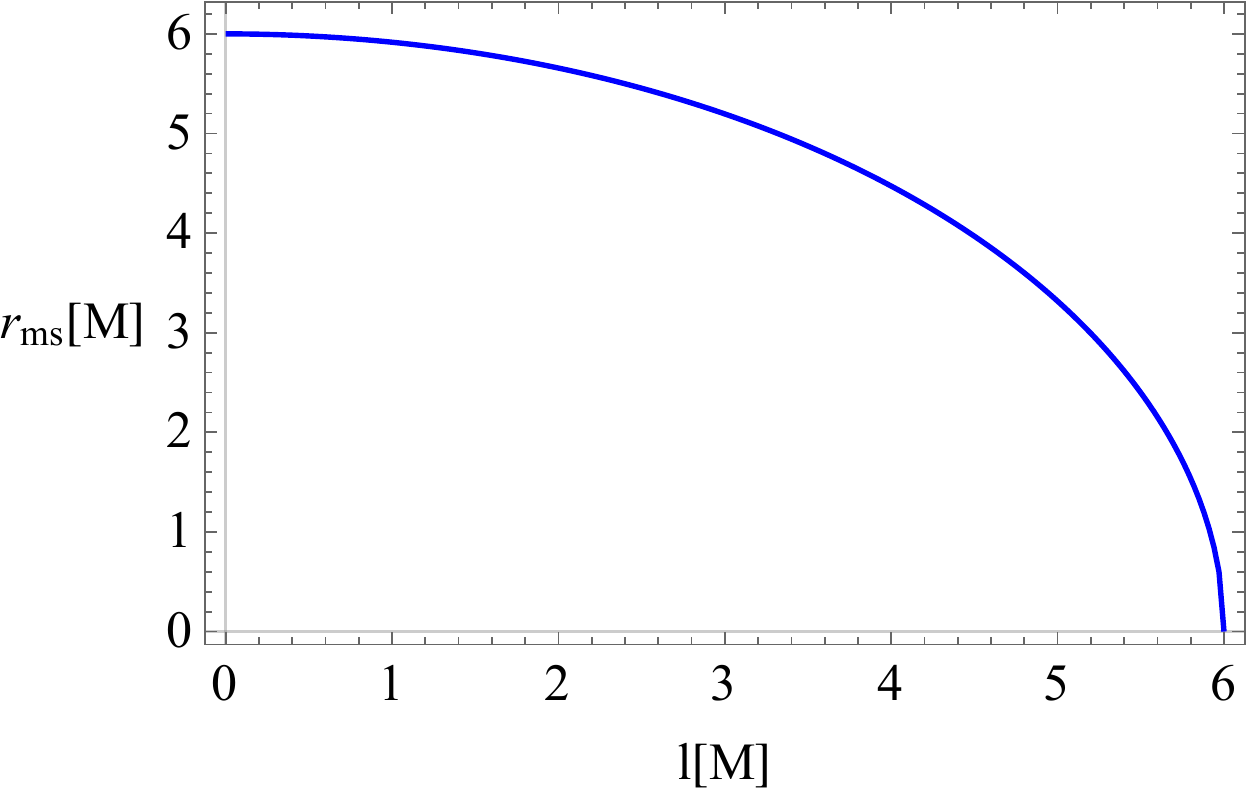}
    \caption{Left panel: The effective potential for different values of $l$ and $M=1$.
 Right panel: The radius of the marginally stable circular orbit or the ISCO for different values of $l$ and $M=1$. }
\end{figure*}

\section{Embedding Diagram}
	Let us study the geometry of the Simpson-Visser metric by embedding it into a higher-dimensional Euclidean space. To simplify the problem let us consider the  equatorial plane $\theta=\pi/2$ at a fixed moment  $t=$ Constant, in that case we have
	\begin{equation}
	ds^2=\frac{dr^2}{1-\frac{b(r)}{r}}+\mathcal{R}^2d\phi^2, \label{emb}
	\end{equation}
	where
	\begin{equation}
	b(r)= r\left(1-f(r)\right),\,\,\,\mathcal{R}^2=r^2+l^2.
	\end{equation}
	Let us embed this black hole metric into three-dimensional Euclidean space  in the cylindrical coordinates,
		\begin{eqnarray}
		ds^2&=&dz^2+d\mathcal{R}^2+\mathcal{R}^2d\phi^2,
		\end{eqnarray}
		which can be written as follows
		\begin{eqnarray}
		ds^2&=&\left[\left(\frac{d\mathcal{R}}{dr}\right)^2+\left(\frac{dz}{dr}\right)^2 \right]dr^2+\mathcal{R}^2d\phi^2.
		\end{eqnarray}
		From these equations,  we find that
	\begin{equation}
	\frac{dz}{dr}=\pm \sqrt{\frac{r}{r-b(r)}-(\frac{d\mathcal{R}}{dr})^2},
	\end{equation}

 As a particular case we will consider the traversable wormhole with $l>2$. Note that the integration of the last expression cannot be accomplished analytically. Invoking numerical techniques allows us to illustrate the embedding diagrams for the Simpson-Visser traversable wormhole given in figure (\ref{geometry}).

\begin{figure*}
    \centering
    \includegraphics[scale=0.50]{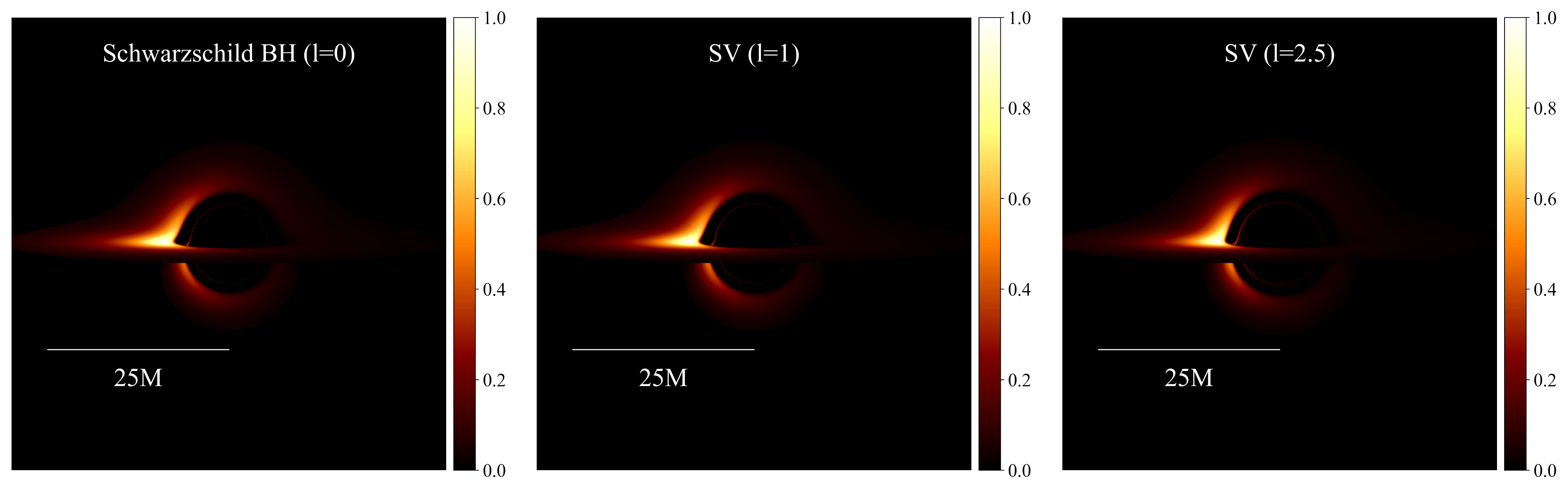}
    \caption{Images of the Novikov-Thorne thin-accretion disk around Schwarzschild Black hole and Simpson-Visser regular Black hole with $l=1$ and wormhole with $l=2.5$ for inclination 90$\degree$.}
    \label{fig:NT}
\end{figure*}

\begin{figure*}
    \centering
    \includegraphics[scale=0.50]{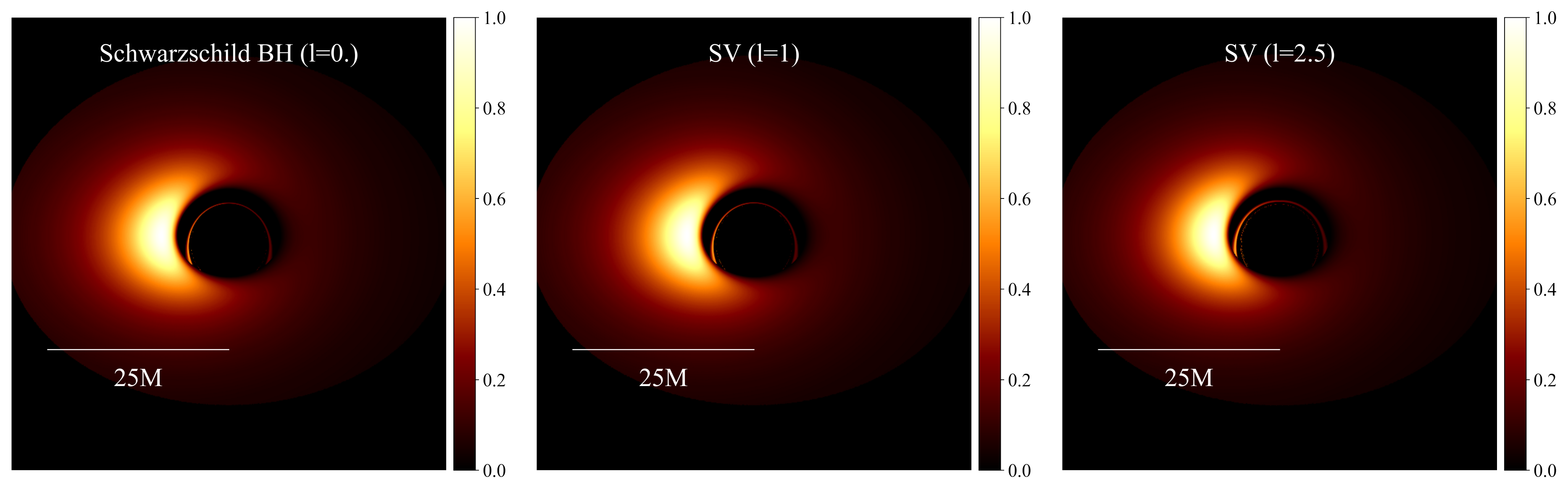}
    \caption{Images of the Novikov-Thorne thin-accretion disk around Schwarzschild Black hole and Simpson-Visser regular Black hole with $l=1$ and wormhole with $l=2.5$ for inclination 45$\degree$.}
    \label{fig:NT2}
\end{figure*}

\section{Novikon-Thorne thin accretion disks}
Massive particles move in stable circular timelike geodesics in the geometrically thin accretion disk. The Novikov-Thorne model of a thin accretion disk is considered here. There are two constants of motion along the timelike geodesics because we are dealing with spherically symmetric and static spacetimes. The particles' energy $(\Tilde{E})$ and angular momentum $(\Tilde{L})$ per unit rest mass, which are represented as,
\begin{equation}
    \Tilde{E}=\left(1-\frac{2M}{\sqrt{r^2+l^2}}\right)\Dot{t}, \; \; \; \; \; \; \;
    \Tilde{L}=(r^2+l^2)\Dot{\phi},
   \label{conserved}
\end{equation}
where over dot represents the derivative with respect to proper time $(\tau)$.
Using the above conserved quantities and timelike geodesic condition $(ds^2=-m^2)$, we can write the general form of spherically symmetric and static spacetime (\ref{SVmetric}) as,
\begin{equation}
    \Tilde{E}^2=\Dot{r}^2+ V_{eff}(r),
\end{equation}
this above equation represents the total energy of the particle, where $V_{eff}(r)$ is the effective potential of the Simpson-Visser spacetime for timelike geodesics. The effective potential of the Simpson-Visser spacetime is defined as,
\begin{equation}
    V_{eff}(r)=\Tilde{L}^2\frac{\left(1-\frac{2M}{\sqrt{r^2+l^2}}\right)}{(r^2+l^2)}+\left(1-\frac{2M}{\sqrt{r^2+l^2}}\right)m^2.
\end{equation}
where, for photons $m = 0$ and for test particles $m > 0$ (see Fig. 2). The sign of the circular geodesic orbits determines their stability. The condition $V''_{eff}(r_c)=0$ determines the marginally innermost stable circular orbit (ISCO). It indicates the ISCO position at the radius,
\begin{eqnarray}
r_{\rm ISCO}=\sqrt{(6M)^2-l^2}.
\label{isco}
\end{eqnarray}
The horizon position of a regular black hole $(l<2M)$ is defined as,
\begin{equation}
    r_h=\sqrt{(2M)^2-l^2}.
\end{equation}
For the shadow size, one can calculate the radius of the circular orbit of a photon $(l<3M, ds^2=0)$ as,
\begin{equation}
    r_{ph}=\sqrt{(3M)^2-l^2},
\end{equation}
and the critical impact parameter $b = E/L$ of the photon circular orbits is given by the expression,
\begin{equation}
    b^2_{ph}=27M^2,
\end{equation}
having the same size as a photon circular orbit around a Schwarzschild black hole and does not depend on $l$ \cite{sv7}. Note that for $l>3M$, there is no photon sphere but we have a contribution from the wormhole throat.
Now, the stable circular timelike geodesic can be obtained by satisfying the conditions,
\begin{equation}
    V_{eff}(r)=\Tilde{E}^2, \; \; \; \; 
    V'_{eff}(r)=0, \; \; \; \; 
     V''_{eff}(r)<0,
\end{equation}
where, the first two conditions are useful to get $\Tilde{E}$ and $\Tilde{L}$ as a function of radius of the circular orbit as,
\begin{equation}
    \Tilde{E}=\frac{1-\frac{2M}{\sqrt{r^2+l^2}}}{\sqrt{1-\frac{2M}{\sqrt{r^2+l^2}}-(r^2+l^2)\Omega^2}},
    \label{Etilta}
\end{equation}
\begin{equation}
    \Tilde{L}=\frac{(r^2+l^2)\sqrt{\frac{M}{(r^2+l^2)^{3/2}}}}{\sqrt{1-\frac{2M}{\sqrt{r^2+l^2}}-(r^2+l^2)\Omega^2}},
    \label{Ltilta}
\end{equation}
where, $\Omega$ gives the angular momentum of the particles in terms of the equation \cite{Shaikh:2019hbm},
\begin{equation}
    \Omega=\frac{d\phi}{dt}=\sqrt{\frac{f'(r)}{2r}}=\sqrt{\frac{M}{(r^2+l^2)^{3/2}}},
    \label{Omega}
\end{equation}
with prime representing a derivative with respect to the radial coordinate `r'. For more details about the properties of the circular geodesic motion see \cite{Younsi:2016azx}.

\begin{figure}
    \centering
        \includegraphics[width=\columnwidth]{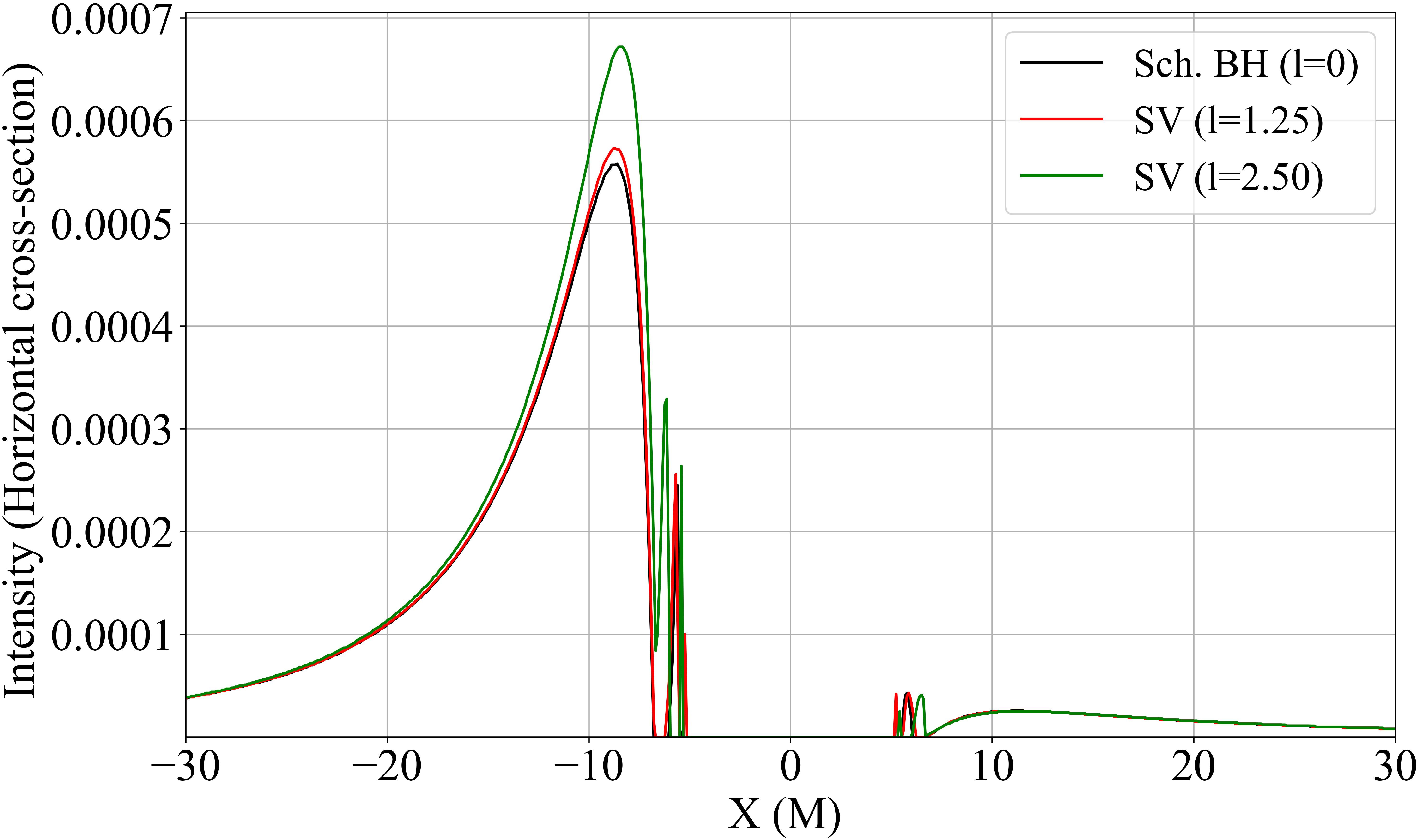}
    \caption{The plot shows the horizontal cross-sectional intensity of the Novikov-Thorne accretion disks for $l=0$ and $l=1.25$, respectively.}
\end{figure}

\begin{figure*}
    \centering
        \includegraphics[scale=0.68]{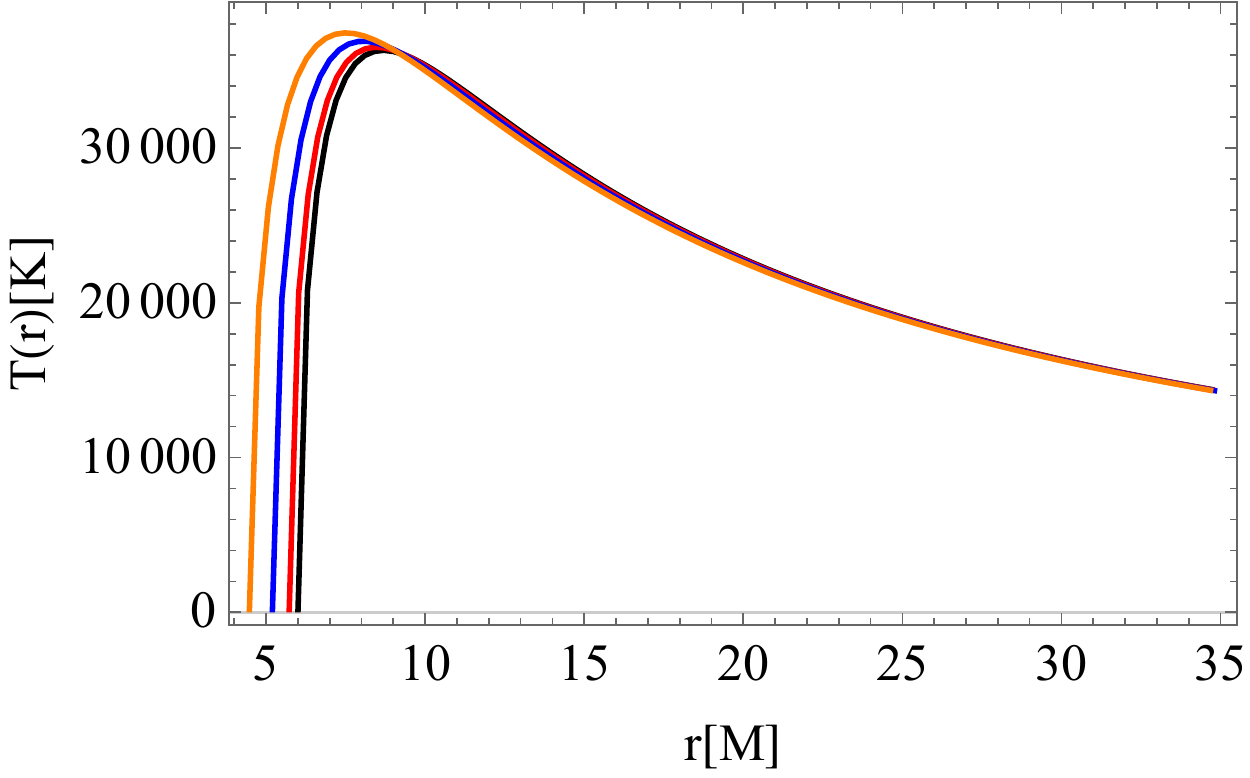}
    \includegraphics[scale=0.70]{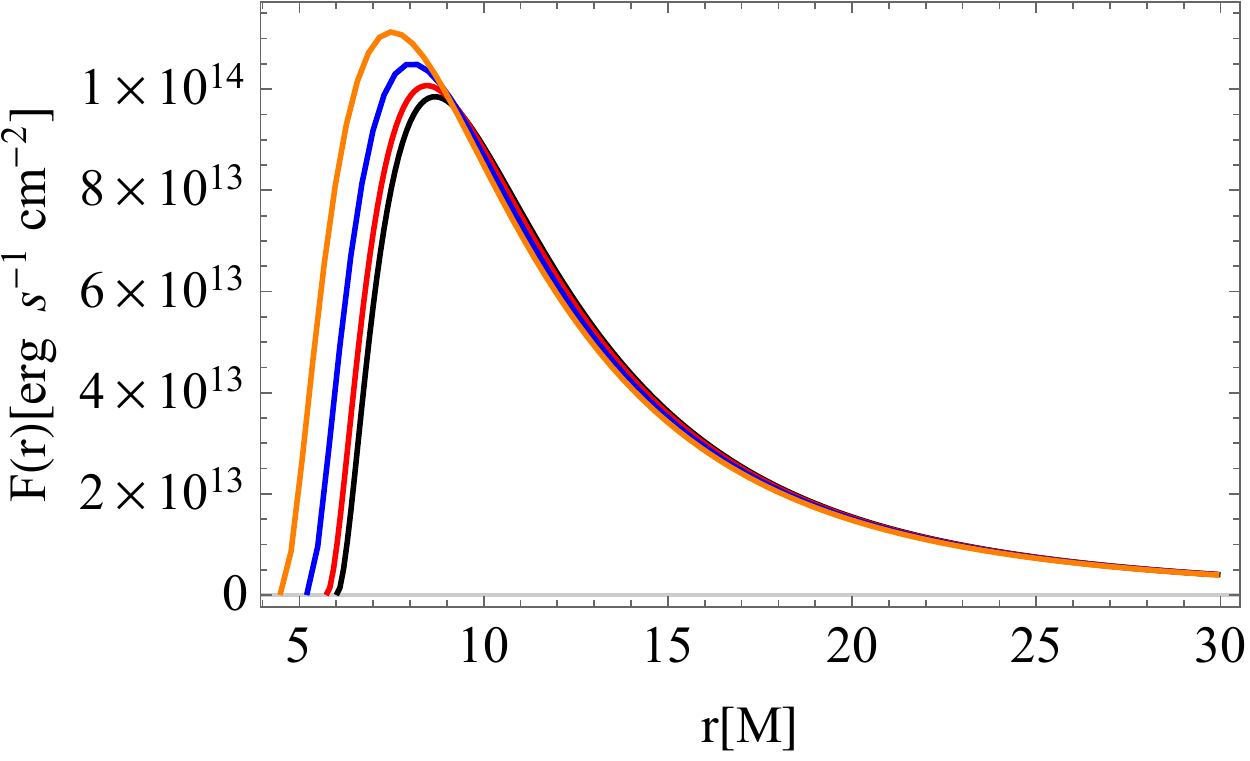}
    \caption{Left panel: The temperature profile of the thin accretion disk for different values of $l$ using the case $M=10^6 M_0$ and $\dot M_0 \sim 2.5\,\times 10^{-5} M_{\odot} /{\rm yr}$. Right panel: The  radiated energy flux over the thin accretion disk for different values of $l$ and $M=1$. In both plots we have $l=0$ (black curve), $l=1.8$ (red curve), $l=3.0$ (blue curve) and $l=4.0$ (orange curve).}
\end{figure*}

\begin{figure*}
    \centering
        \includegraphics[scale=0.65]{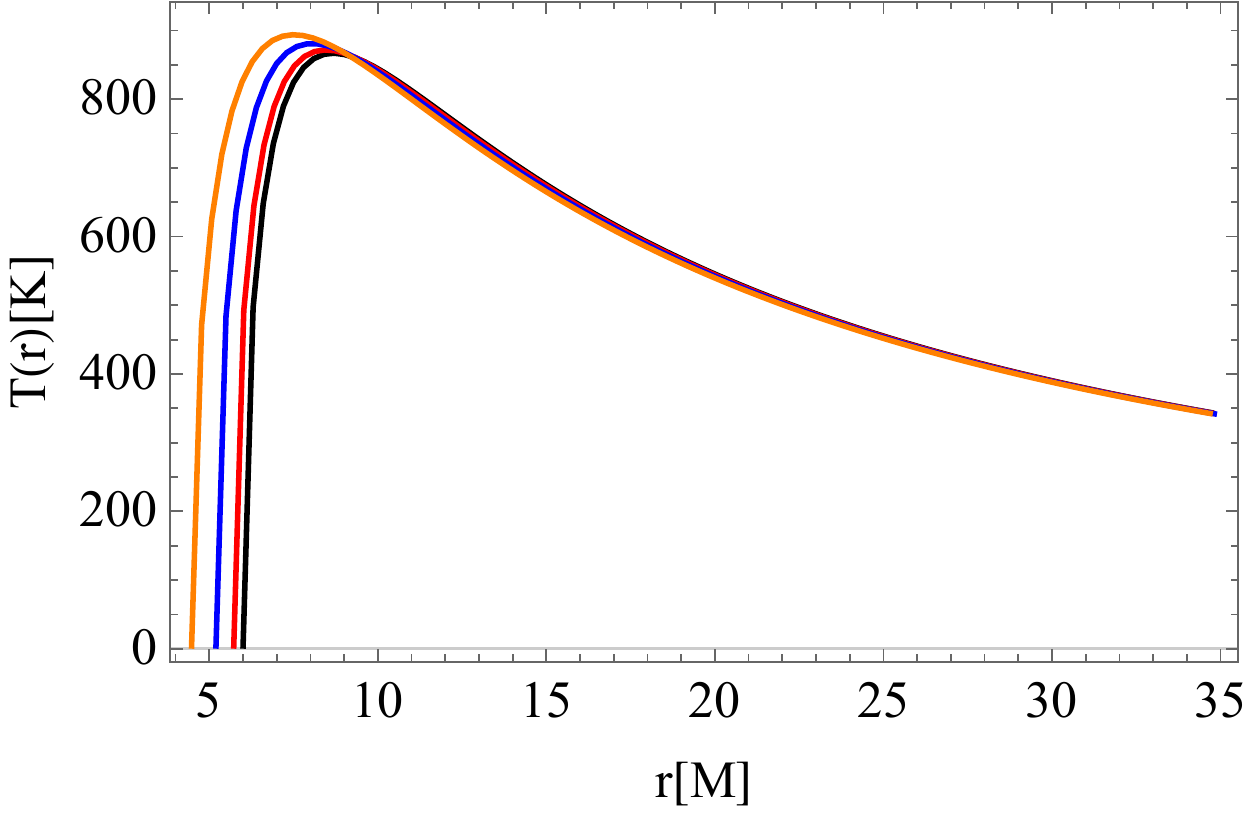}
    \includegraphics[scale=0.72]{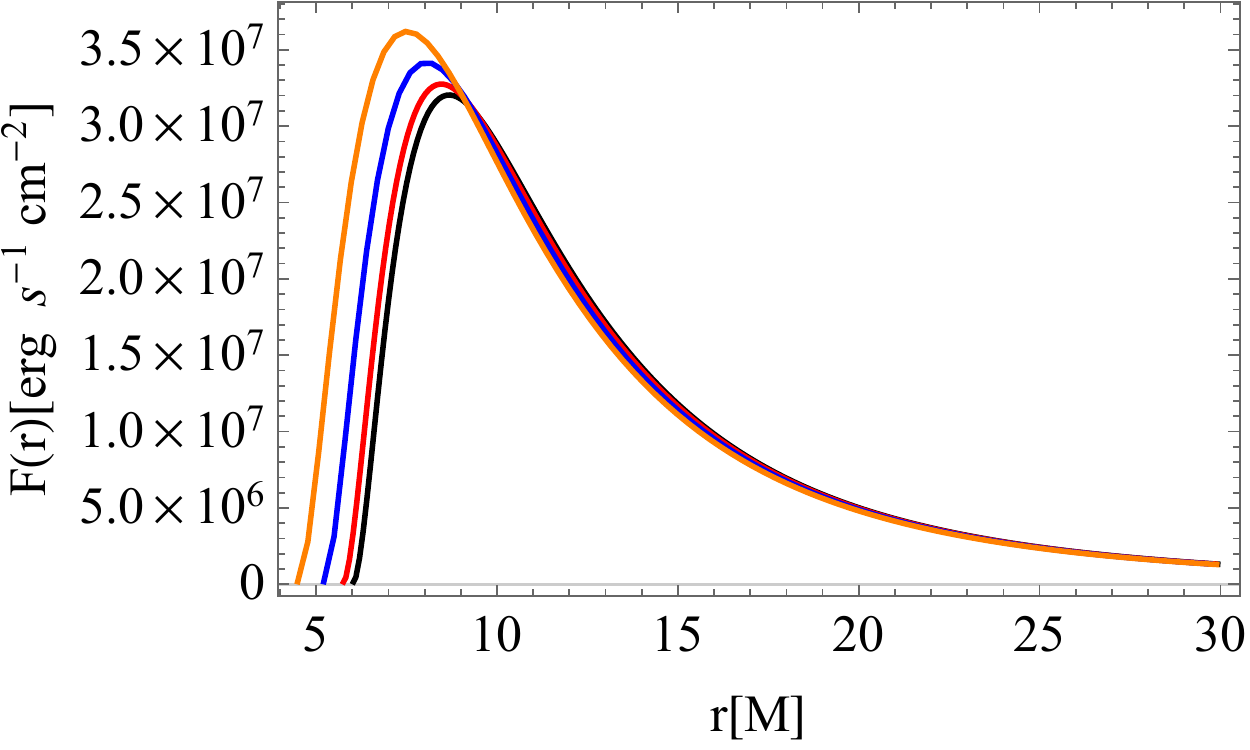}
    \caption{Left panel: The temperature profile of the thin accretion disk for different values of $l$ using the case $M=10^6 M_0$ and $\dot M_0 \sim 8\,\times 10^{-12} M_{\odot} /{\rm yr}$. Right panel: The  radiated energy flux over the thin accretion disk for different values of $l$ and $M=1$. In both plots we have $l=0$ (black curve), $l=1.8$ (red curve), $l=3.0$ (blue curve) and $l=4.0$ (orange curve) in both plots.}
\end{figure*}

\begin{figure*}
    \centering
            \includegraphics[scale=0.69]{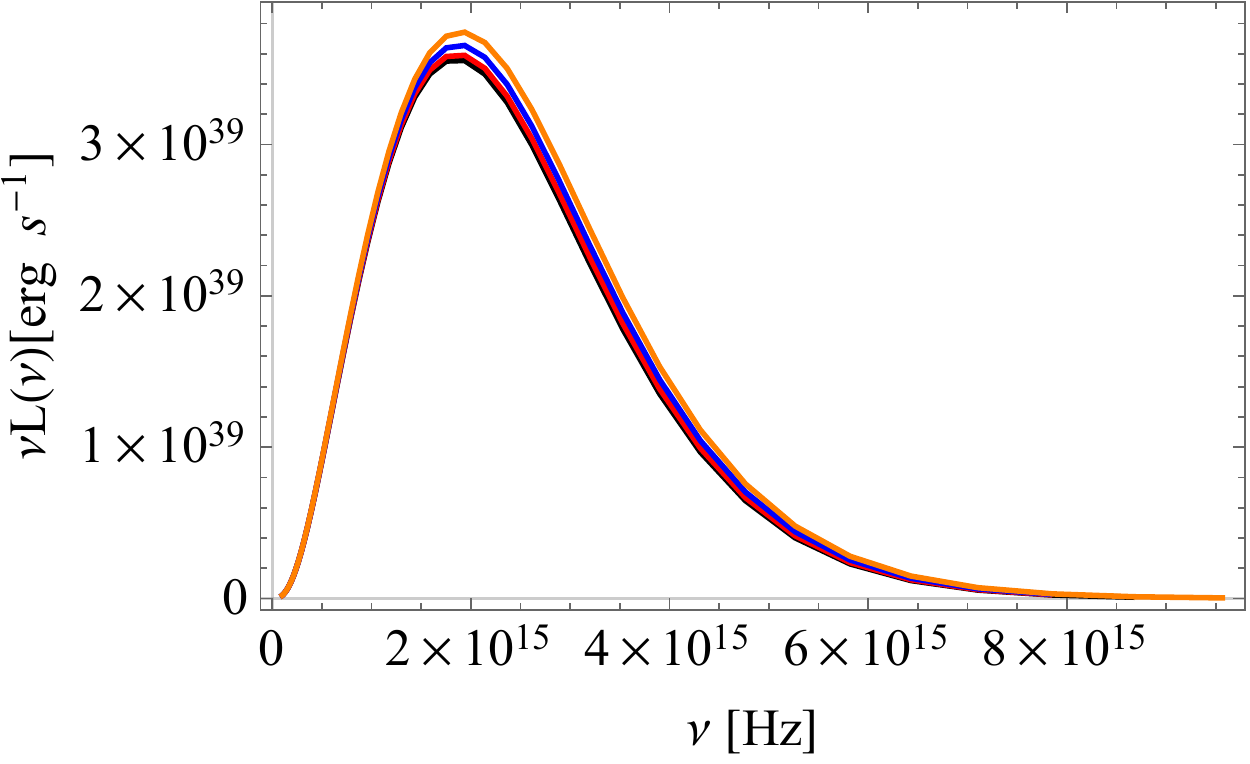}
            \includegraphics[scale=0.71]{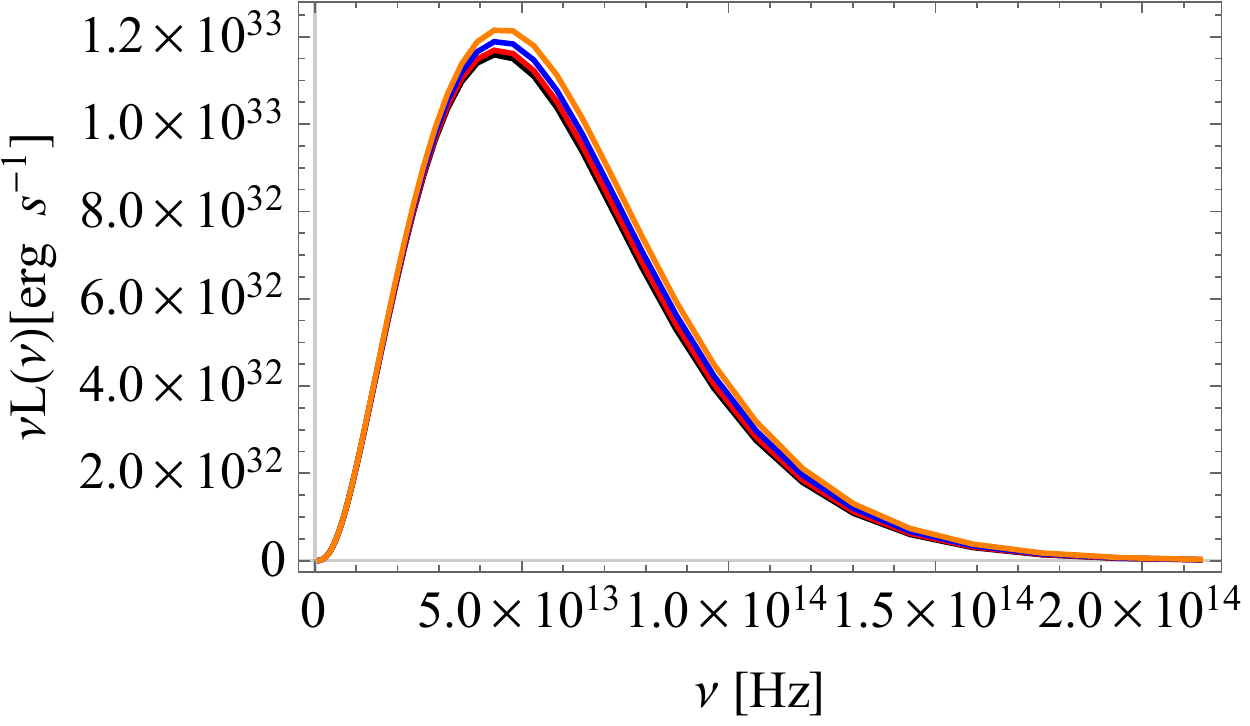}
    \caption{Left panel: The observed luminosity of the thin accretion disk for different values of $l$ using the case $M=10^6 M_0$ and $\dot M_0 \sim 2.5\,\times 10^{-5} M_{\odot} /{\rm yr}$.  Right panel: The observed luminosity of the thin accretion disk for different values of $l$ using the case $M=10^6 M_0$ and $\dot M_0 \sim 8\,\times 10^{-12} M_{\odot} /{\rm yr}$. We have $l=0$ (black curve), $l=1.8$ (red curve), $l=3.0$ (blue curve) and $l=4.0$ (orange curve) in both plots. }
\end{figure*}

In this part, we use the Novikov-Thorne model of a thin accretion disk composed of anisotropic fluid moving in the equatorial plane to examine the steady-state thin accretion disk. Certain structure equations govern the physical properties of the accretion disk, which arises from the necessity of the fluid's rest mass, energy, and angular momentum conservation. In this setup, the accreting matter in the disk can be described by the stress-energy tensor given by  \cite{1}.
\begin{eqnarray}
T^{\mu\nu}=\rho_0 u^\mu u^\nu +2 u^{(\mu}q^{\nu)}+t^{\mu\nu},
\end{eqnarray}
where
\begin{eqnarray}
u_\mu q^\mu&=&0,\\
u_\mu t^{\mu\nu}&=&0.
\end{eqnarray}
In the above equations we have the introduced the following quantities: $\rho_0$, $q^\mu$, and $t^{\mu\nu}$, which are known as the rest mass density of the accreting matter, the energy flow vector, and the stress tensor, respectively. These quantities are specified in the averaged rest-frame of the particle described by the 4-velocity $u^\mu$. If the rest mass is conserved, the equation can be used as,
\begin{equation}
\nabla_\mu (\rho_0 u^\mu)=0
\end{equation}
As a result, the disk radius has no effect on the time averaged rate of rest mass accretion.
\bqn
\dot{M}_0 &=& -2\pi \sqrt{-g} \Sigma u^r={\rm const.},
\eqn
where $\Sigma(r)$ denotes the time-averaged surface density in cylindrical coordinates, and
\bqn
\Sigma (r) &=& \int^H_{-H}<\rho_0>dz,
\eqn

The law of energy conservation and the law of angular momentum conservation both state that
\begin{eqnarray}
\nabla_\mu T^{t \mu}&=&0,\\
\nabla_\mu T^{\phi \mu}&=&0.
\end{eqnarray}

These relationships can be used to obtain the thin disk's time-averaged radial structure equations,
\bqn
&&[\dot{M}_0\tilde{E}-2\pi \sqrt{-g} \Omega W^r_\phi]_{,r} =4 \pi r F(r)\tilde{E}, \\
\eqn
and 
\bqn
&&~~[\dot{M}_0\tilde{L}-2\pi \sqrt{-g} W^r_\phi]_{,r}=4 \pi r F(r) \tilde{L}.
\eqn
where $W^r_\phi$ is the averaged torque and is given by
\bqn
W^r_\phi=\int^H_{-H}<t^r_\phi>dz.
\eqn
At this point, it's important to emphasise that the $\phi$-$r$ component of the stress tensor averaged over a characteristic time scale $\Delta t$  and the azimuthal angle $\Delta \phi=2 \pi$ is the quantity $<t^r_\phi>$. The flux $F(r)$ of the radiant energy over the disk can be represented in terms of the specific energy, angular momentum, and angular velocity of the orbiting particle in the thin accretion by using the energy-angular momentum relation $\tilde{E}_{,r}=\omega \tilde{L}_{,r}$. The flux of electromagnetic radiation emitted from a radial position 'r' of a disk is given by the standard formula, which may be deduced from the equations (\ref{Etilta}),(\ref{Ltilta}), and (\ref{Omega}) \cite{Shaikh:2019hbm},
\begin{equation}
    F(r)=-\frac{\Dot{M}}{4\pi\sqrt{-g}}\frac{\Omega'}{(\Tilde{E}-\Omega\Tilde{L})^2}\int_{r_{in}}^{r} (\Tilde{E}-\Omega\Tilde{L})\Tilde{L'} \,dr ,
\end{equation}
Note that $r_{in}$ represent the inner edge of the disk and $\Dot{M}$ is the mass accretion rate. In Figs. 3 and 4, we show the optical appearance of the Simpson-Visser spacetime surrounded by NT accretion disk. We use the ray-tracing formalism described in \cite{Younsi:2016azx} to numerically integrate the geodesic equation using \texttt{RK45} method with adaptive step size. In Fig. 5, we show the  horizontal cross-sectional intensity for different values of $l$ and highlights the peak intensity for the same.
One can see that, the images are almost similar compared to the Schwarzschild geometry, which makes it difficult to distinguish these two spacetime geometries. However, in what follows we shall study in more details this problem, in particular we shall study in the electromagnetic properties of the Simpson-Visser spacetime and compere to the Schwarzschild geometry. 

\section{The Radiating flux, temperature profile, and the luminosity of thin accretion disk}
In this section, we turn our attention to explore in more details the electromagnetic properties of the accretion disk, such as the radiation flux $F(r)$, the temperature profile, luminosity, and the accretion efficiency. Toward this purpose, we will numerically analyse the behaviour of the above quantities with respect to the parameter $l$. Furthermore we assume a central compact object with mass $M=10^6 M_{\odot}$ (a supermassive black hole), along two cases for the accretion rate given as follows \cite{1,6}
\begin{equation}
\dot M_0 \sim 10^{-5} M_{\odot} /{\rm yr}\,\, \text{and} \,\,\dot M_0 \sim 10^{-12} M_{\odot} /{\rm yr}
\end{equation}

Recall that the radiation flux $F(r)$ given by Eq. (4.24) represent the emitted flux by the thin accretion disk in the Simpson-Visser spacetime. Moreover, we expect that the Stefan-Boltzmann law holds and, therefore, in terms of the effective temperature associated to the accretion disk we have         
\bqn
T_{\rm eff}(r) = \left(\frac{F(r)}{\sigma}\right)^{1/4},
\eqn
where for the Stefan-Boltzmann constant we have $\sigma = 5.67 \times 10^{-5}\; {\rm erg}\; s^{-1} \;{\rm cm}^{-2}\; K^{-4}$. Our analyses shows that, the temperature and the radiated energy flux of the disk increases by increasing the parameter $l$, as can be seen from Figs. 6 and 7. As a specific example, take the case $\dot M_0 \sim 8\,\times 10^{-12} M_{\odot} /{\rm yr}$, we find that the maximal value for the temperature at a particular $l$ occurs at:
\begin{equation}\notag
l=0.0:\,T_{max}\simeq 867.01\, {\rm K}, \,\,r=8.70[M]
\end{equation}
\begin{equation}\notag
l=1.8:\,T_{max}\simeq 871.79 \, {\rm K},\,\,r=8.42 [M]
\end{equation}
\begin{equation}\notag
l=3.0:\,T_{max}\simeq 881.04\, {\rm K}\,\,r=8.09 [M]
\end{equation}

On the other hand, the maximal value for the radiating flux at particular $l$ occurs at:
\begin{equation}\notag
l=0.0:\,F_{max}\simeq 3.203 \times 10^7 {\rm erg}\; s^{-1} \;{\rm cm}^{-2}, \,r=8.70[M]
\end{equation}
\begin{equation}\notag
l=1.8:\,F_{max}\simeq 3.275 \times 10^7 {\rm erg}\; s^{-1} \;{\rm cm}^{-2},\,r=8.42 [M]
\end{equation}
\begin{equation}\notag
l=3.0:\,F_{max}\simeq 3.416 \times 10^7 {\rm erg}\; s^{-1} \;{\rm cm}^{-2},\,r=8.09 [M]
\end{equation}

The observed luminosity $L(\nu)$ of the thin accretion disk around the Simpson-Visser spacetime has a red-shifted black body spectrum while we assume the radiation emitted by the thin accretion disk surface to be perfect black body radiation \cite{1},
\bqn
L(\nu) &=& 4 \pi d^2 I(\nu) 
\eqn
or 
\bqn
L(\nu)&=& \frac{8 \pi h \cos i}{c^2} \int_{r_i}^{r_f} \int_{0}^{2\pi}  \frac{\nu_e^3 r d \phi dr}{\exp{(\frac{h \nu_e}{k_{\rm B} T})}-1},
\eqn
in this equation $i$ is the thin accretion disk's inclination angle around the black hole while $d$ is the distance between the observer and the centre of the thin accretion disk, along with the Planck constant $h$, emission frequency $\nu_e$, as well as the Boltzmann constant $k_{\rm B}$. Furthermore we need to the red-shift factor which can be found from
\bqn
g=\frac{\nu}{\nu_e}=\frac{k_\mu u^\mu_o}{k_\mu u^\mu_e},
\eqn
where $\nu$ is the radiation frequency in the faraway observer's local rest frame. We also have the observer's four-velocity,
\begin{eqnarray}
u^\mu_o=\left(1,0,0,0\right),
\end{eqnarray}
and the the 4-velocity of the emitter
\begin{eqnarray}
u^\mu_e=\left(u^t_e,0,0,\Omega u^t_e\right).
\end{eqnarray}
We can take $r_i=r_{\rm ms}$ and $r_f$ to any arbitrary large distance from the compact object. To illustrate the effect, and see how the parameter $l$ affects the emission spectrum, we calculate the radiation spectrum $\nu L(\nu)$ numerically. We find that with the increase of the parameter $l$ for both accretion rates the observed luminosity increases as shown in Fig. 8. If we take the accretion case dot $M_0 \sim 8\,\times 10^{-5} M_{\odot} /{\rm yr}$, we find that the maximal value for the luminosity at particular $l$ occurs at the frequency $\nu=4.33 \times 10^{13}\, {\rm Hz}$ along with the following values for the luminosity:
\begin{equation}\notag
l=0.0:\,\nu L_{max}\simeq 1.158 \times 10^{33} {\rm erg}\; s^{-1}, 
\end{equation}
\begin{equation}\notag
l=1.8:\,\nu L_{max}\simeq 1.169 \times 10^{33} {\rm erg}\; s^{-1},
\end{equation}
\begin{equation}\notag
l=3.0:\,\nu L_{max}\simeq 1.188 \times 10^{33} {\rm erg}\; s^{-1}.
\end{equation}

Finally, we can think about accretion efficiency, which is calculated as the ratio of the rate of photons escaping from the disk surface to infinity compared to the rate of mass-energy transport to the black hole. There is a simple way to estimate the efficiency $\epsilon$ of the accretion disk, since it can be found to be proportional to the specific energy of the moving particle in the disk, as measured at the marginally stable orbit by
\bqn
\epsilon = 1- \tilde E_{\rm ms}.
\eqn

%\begin{figure}
%    \centering
%    \includegraphics[scale=0.63]{eps1.pdf}
%    \caption{The accretion efficiency $\epsilon$ which is shown to be constant and does not depend on $l$ for $M=1$.}
%\end{figure}

Interestingly, for the efficiency of the accretion disk in the Simpson-Visser spacetime we find that this quantity is not affected by the parameter $l$ and it is given by
\bqn
\epsilon = \frac{3-2 \sqrt{2}}{3}\simeq 0.0571909588.
\eqn
Although the radiating energy changes, we found that based on the accretion efficiency $\epsilon$ one cannot distinguish the Simpson-Visser metric from the Schwarzschild geometry. This quantity is important and provides information about the efficient engine, in our case the accretion disk, for converting accreting matter's energy into electromagnetic radiation. Its interesting to see that this quantity remains constant and does not depend on the parameter $l$. From the plot of luminosity as a function of the frequency of the emitted electromagnetic waves from the accretion disk we saw that it depends on the spacetime geometry. This in turn, allows to distinguish different spacetime geometries.  In addition, the observed luminosity depends on the accretion mass rate. That means, in order to constrain a given solution, we need a precise value for the accretion mass rate for a given source and observational data for the luminosity which may be obtained by means of other astrophysical observations (here, we have used two examples of the mass accretion rate). With this information in hand, we can compare and test different geometries. Another quantity of interest is the radiation energy flux, which also encodes information about the spacetime geometry. In this direction, we point out a similar method based on the thermal radiation used by Bambi \cite{t1} to analyse the photon flux number density as measured by a distant observer to test the nature of the black hole and modified gravity.  In addition, the so called continuum-fitting method has been used for example in Ref. \cite{t2} to constrain different geometries.
\section{Conclusions}
The Simpson-Visser model has the additional benefit of analytically smoothly shifting from black holes (singular/regular) to wormholes (one-way/traversable). In some ways, the regular black hole shown above in this paper is in fact, the conventional Schwarzschild solution for $l = 0$. This is a clear contrast to the scenario in which a collapsing regular black hole “bounces” back into our own universe, and it is a situation worth considering in its own right.  It'll be interesting to see if we can tell the difference between a regular and singular black hole and a wormhole. This metric is particularly interesting because, depending on the value of the regularisation parameter $l$, it can either describe a regular/singular black hole or a traversable/one-way wormhole.
Therefore, in this paper, we studied the optical appearance of a thin accretion disk in the Simpson-Visser spacetime.  We calculate and illustrate the observed flux distributions as viewed by a distant observer. The image generated in the Simpson-Visser metric looks similar to that of Schwarzschild. In addition, we studied the electromagnetic properties of the accretion disk such as the temperature and the radiation flux of the energy by the accretion disk. It is shown that by increasing the regularized parameter $l$, the temperature, as well as the radiating flux increases, compared to the Schwarzschild geometry. Thus, for these solutions, we conclude that
the specific signatures, that appear in the electromagnetic spectrum, lead to the possibility of distinguishing
wormhole geometries from the Schwarzschild solution by
using astrophysical observations of the emission spectra
from accretion disks. As an interesting result, we show that the accretion efficiency remains constant and does not depend on $l$. In the near future, we plan to use astrophysical data to constrain the regularized parameter $l$.

\end{document}